\documentclass[preprint,aps]{revtex4}
\usepackage{mathrsfs}

\usepackage{graphicx}
\usepackage{multirow}
\usepackage{makecell}
\usepackage{booktabs}

\begin{document}

\title{Emergence of Superconductivity from Fully Incoherent Normal State in an Iron-Based Superconductor (Ba$_{0.6}$K$_{0.4}$)Fe$_2$As$_2$}

 \author{Jianwei Huang$^{1,2}$, Lin Zhao$^{1,*}$, Cong Li$^{1,2}$, Qiang Gao$^{1,2}$, Jing Liu$^{1,2}$, Yong Hu$^{1,2}$, Yu Xu$^{1,2}$, Yongqing Cai$^{1,2}$, Dingsong Wu$^{1,2}$, Ying Ding$^{1,2}$, Cheng Hu$^{1,2}$, Huaxue Zhou$^{1}$, Xiaoli Dong$^{1,2}$, Guodong Liu$^{1}$, Qingyan Wang$^{1}$, Shenjin Zhang$^{3}$, Zhimin Wang$^{3}$, Fengfeng Zhang$^{3}$, Feng Yang$^{3}$, Qinjun Peng$^{3}$, Zuyan Xu$^{3}$, Chuangtian Chen$^{3}$ and Xingjiang Zhou$^{1,2,4,5,*}$}

\affiliation{
\\$^{1}$Beijing National Laboratory for Condensed Matter Physics, Institute of Physics, Chinese Academy of Sciences, Beijing 100190, China.
\\$^{2}$University of Chinese Academy of Sciences, Beijing 100049, China.
\\$^{3}$Technical Institute of Physics and Chemistry, Chinese Academy of Sciences, Beijing 100190, China.
\\$^{4}$Songshan Lake Materials Laboratory, Dongguan, Guangdong 523808, China.
\\$^{5}$Collaborative Innovation Center of Quantum Matter, Beijing 100871, China.
}

\date{November 1, 2018}

\maketitle

{\bf In unconventional superconductors, it is generally believed that understanding the physical properties of the normal state is a pre-requisite for understanding the superconductivity mechanism. In conventional superconductors like niobium or lead, the normal state is a Fermi liquid with a well-defined Fermi surface and well-defined quasipartcles along the Fermi surface.  Superconductivity is realized in this case by the Fermi surface instability in the superconducting state and the formation and condensation of the electron pairs (Cooper pairing). The high temperature cuprate superconductors, on the other hand, represent another extreme case that superconductivity can be realized in the underdoped region where there is neither well-defined Fermi surface due to the pseudogap formation nor quasiparticles near the antinodal regions in the normal state.  Here we report a novel  scenario that superconductivity is realized in a system with well-defined Fermi surface but without quasiparticles along the Fermi surface in the normal state. High resolution laser-based angle-resolved photoemission measurements have been performed on an optimally-doped iron-based superconductor (Ba$_{0.6}$K$_{0.4}$)Fe$_2$As$_2$.  We find that, while sharp superconducting coherence peaks emerge in the superconducting state on the hole-like Fermi surface sheets, no quasiparticle peak is present in the normal state. Its electronic behaviours deviate strongly from a Fermi liquid system.  The superconducting gap of such a system exhibits an unusual temperature dependence that  it is nearly a constant in the superconducting state and abruptly closes at T$_c$.  These observations have provided a new platform to study unconventional superconductivity in a non-Fermi liquid system. }\\

{\bf keywords: (Ba$_{0.6}$K$_{0.4}$)Fe$_2$As$_2$, ARPES, Superconducting Gap, Normal State, Non-Fermi Liquid}\\

\vspace{3mm}

\section{Introduction}

It is a general consensus that understanding of the normal state is a pre-requisite for understanding the superconducting state. In conventional superconductors, superconductivity emerges from the normal state where the low energy excitations can be well described by the Fermi liquid theory. There are well-defined quasiparticles that form well-defined Fermi surface. Superconductivity is realized by the instability of the Fermi surface and formation of electron Cooper pairs along the Fermi surface\cite{JBardeen1957JRSchrieffer}.   For high temperature cuprate superconductors, great efforts have been paid to understand the unusual normal state and it is generally believed that it is a pre-requisite to understand its superconducting state and high temperature superconductivity mechanism\cite{CMVarma1989SSchmittRink,PWAnderson1992Science,VJEmery1995SAKivelson,AJSchofield1999ContemPhys,PatrickALee2006XGWen,PWAnderson2007Science,KHaule2009GKotliar,2017LedererSamuel_PNAS}. In cuprate superconductors, particularly in the underdoped region, there is no well-defined Fermi surface formed near the antinodal region and Fermi arc or Fermi pockets are formed due to the occurence of the pseudogap in the normal state\cite{HDing1996JGiapintzakis,AGLoeser1996AKapitulnik}. The electronic state near the antinodal region is highly incoherent in the normal state without formation of quasiparticles. Such an extreme normal state electronic structure, with neither well-defined Fermi surface nor quasiparticles,  has posed a challenge in understanding its underlying origin and the emergence of superconductivity from such an unusual normal state\cite{MRNorman1998JCCampuzano,JCCampuzano2006AKanigel,XJZhou2009JMeng,DSDessau2012TJReber}. Between the two extreme cases of conventional superconductors and cuprate superconductors, it is intriguing to ask whether it is possible to have a scenario that superconductivity can be realized in a system that has well-defined Fermi surface but without quasiparticles in the normal state, and if there is such a case, how about its superconducting properties and superconductivity mechanism.

The iron-based superconductors show some similarities with cuprates, and many discussions have been focused on their non-Fermi liquid normal state\cite{MAichhorn2010MImada,SKasahara2010YMatsuda,Ryu2013QMSi,MNakajima2013SUchida,JGAnalytis2014IRFisher}. However,  decisive evidence of the non-Fermi liquid normal state behaviors is still lacking.  Here we report clear electronic evidence of non-Fermi liquid behaviors in an iron-based superconductor and establish a new superconducting system where there is a well-defined Fermi surface but no quasiparticles in the normal state. We have carried out high resolution laser-based angle-resolved photoemission  (ARPES) measurements on a prototypical iron-based superconductor, the optimally-doped (Ba$_{0.6}$K$_{0.4}$)Fe$_2$As$_2$. We find that, while sharp superconducting coherence peaks emerge in the superconducting state, the normal state becomes fully incoherent with no quasiparticles along a well-defined Fermi surface. The normal state  exhibits clear non-Fermi liquid electronic behaviors.  The superconducting gap exhibits an unusual behavior that it is nearly constant in the entire superconducting state but drops abruptly to zero at the superconducting transition temperature; this is inconsistent with the conventional BCS picture. Our results have provided electronic evidence in identifying a non-Fermi liquid scenario in the normal state of the iron-based superconductors. They will provide key insights in understanding the emergence of coherent superconducting state out of the fully incoherent normal state in unconventional superconductors.

\section{Experimental}
(Ba$_{0.6}$K$_{0.4}$)Fe$_2$As$_2$ single crystals were grown using a FeAs flux method with excess FeAs as flux\cite{GFChen2008JLLuo}.   The plate-like single crystals with size up to 5mm$\ast$5mm were obtained. The samples were characterized by electrical resistivity, magnetic susceptibility and specific heat measurements. The resistivity in Fig. 1a exhibits a nearly linear temperature dependence between the superconducting transition temperature (T$_c$)  and 150 K which is similar to  that observed in the optimally-doped cuprate superconductors\cite{ATFiory1987MGurvitch,GSBoebinger2000SOno}. The onset of  superconducting transition in resistivity is at $\sim$39 K.  Extrapolation of  the resistivity between 40 K and 65 K temperature range gives a negative value (linear extrapolation) or nearly zero (linear plus parabolic extrapolation) at zero temperature (see inset of Fig. 1a) which indicates weak impurity scattering in the grown single crystals.  The magnetic susceptibility measurement exhibits a superconducting transition onset temperature at 38.5 K with a narrow transition width of $\sim$0.5 K (10$\%$-90$\%$ transition width)  (Fig. 1b).  In particular, not only the ZFC (zero field cooling) curve shows a nearly 100$\%$ shielding fraction, the FC (field cooling) curve also shows more than 20$\%$ drop in the superconducting state that is much higher than that reported before. In the specific heat measurement (Fig. 1c),  the value of ${\Delta}C_e/\gamma_nT_c$ is about 2 which indicates (Ba$_{0.6}$K$_{0.4}$)Fe$_2$As$_2$ is in a strong-coupling region. The large specific heat jump at T$_c$ (top-left inset in Fig. 1c) and the linear extrapolation to 0 at T = 0 K for the data plotted vs T$^{2}$ (bottom-right inset in Fig. 1c), together with the above resistivity and magnetic measurement results, strongly indicate the high quality of our grown (Ba$_{0.6}$K$_{0.4}$)Fe$_2$As$_2$ single crystal samples.

High resolution angle-resolved photoemission measurements were performed using a new lab-based ARPES system equipped with the 6.994 eV vacuum-ultra-violet (VUV) laser and the angle-resolved time-of-flight electron energy analyzer (ARToF) with the capability of simultaneous two-dimensional momentum space detection\cite{XJZhou2016CLWang,XJZhou2018LaserARPES}.   One advantage of the ARToF analyser is that it has much weaker non-linearity effect so that the measured signal is intrinsic to the sample. The energy resolution was set at 1 meV and the angular resolution was $\sim$0.3$^o$ corresponding to 0.004 ${\AA}^{-1}$ momentum resolution at the photon energy of 6.994 eV. All the samples were cleaved \emph{in situ} at low temperature of 30 K and measured in ultrahigh vacuum with a base pressure better than 5$\times$10$^{-11}$ mbar. For each measurement, sample temperature is precisely controlled to be within $\pm$0.5 K accuracy.  The measurements have been performed on many samples and the results are fully reproducible.  The Fermi level is referenced by measuring on clean polycrystalline gold that is electrically connected to the sample or checked by the Fermi level of the measured sample in the normal state.

\section{Results and Discussions}
Figure 1d and 1e show the Fermi surface mapping of the optimally-doped (Ba$_{0.6}$K$_{0.4}$)Fe$_2$As$_2$ superconductor  (T$_c$$\sim$38.5 K), measured at 13 K,  obtained by integrating the spectral weight within an $\pm$ 5 meV energy window relative to the Fermi level E$_F$.  Two independent areas of the momentum space are covered; in each area, all the data points are measured under the same experimental condition thanks to the capability of simultaneous two-dimensional momentum detection of our laser ARToF ARPES system.  From Fig. 1d and 1e,  two hole-like Fermi surface sheets around $\Gamma$ can be clearly identified. The inner hole pocket is labelled as $\alpha$ while the outer one is labelled as $\gamma$ in Fig. 1d and 1e.  By fitting the momentum distribution curves (MDCs) across the two Fermi surface sheets, the Fermi momentum k$_F$s are quantitatively determined,  as shown in Fig. 1f. Fig. 1g and 1h show symmetrized energy distribution curves (EDCs) along the inner hole pocket (Fig. 1g) and the outer hole pocket (Fig. 1h).  Sharp superconducting coherence peaks are clearly observed along both hole pockets. It is a standard procedure to symmetrize EDCs on the Fermi surface and fit them with a phenomenological model\cite{MRNorman1998JCCampuzano} to extract the superconducting gap size and the single-particle scattering rate using the self-energy $\Sigma(k_F,\omega) =  -i\Gamma_1 + \Delta^2/(\omega + i0^+) $ where $\Gamma_1$ represents single-particle scattering rate and $\Delta$ represents superconducting gap. The symmetrized EDCs can be well fitted by such a formula, as shown in Fig. 1g and 1h, and the extracted superconducting gap and the single-particle scattering rate are shown in Fig. 1i and 1j, respectively, for both measurements in Fig. 1d and 1e. The superconducting gap around the entire inner hole-pocket is plotted in Fig. S1 (see Supplementary Materials).

The superconducting gap of (Ba$_{0.6}$K$_{0.4}$)Fe$_2$As$_2$ superconductor (Fig. 1i) shows a clear Fermi surface-dependence.  The inner  hole pocket $\alpha$ has a larger superconducting gap of $\sim$9.3 meV while the outer hole pocket $\gamma$ shows a smaller gap of $\sim$5.0 meV. Moreover, with a much improved experimental precision, we find that the measured superconducting gaps in (Ba$_{0.6}$K$_{0.4}$)Fe$_2$As$_2$ superconductor are extremely isotropic. The gap anisotropy (defined by the maximum deviation of the gap size divided by the average value) is less than 2$\%$ for the inner hole pocket  (Fig. 1i and Fig. S1) and less than 4$\%$ for the outer hole pocket (Fig. 1i).  The gap function exhibits a slight deviation from $|cos(k_x) cos(k_y)|$ formula as shown in the inset of Fig. 1i.  The single-particle scattering rate  is also isotropic along the two Fermi surface sheets, as seen in Fig. 1j. Interestingly,  the superconducting gap size is proportional to the single-particle scattering rate (inset of Fig. 1j): the inner $\alpha$ hole pocket shows a larger single-particle scattering rate that also has a larger superconducting gap.

Our present results have resolved a long-standing controversy regarding the superconducting gap in the optimally-doped  (Ba$_{0.6}$K$_{0.4}$)Fe$_2$As$_2$ superconductor. Fermi surface-dependent superconducting gap was reported by many groups in optimally-doped (Ba$_{0.6}$K$_{0.4}$)Fe$_2$As$_2$ superconductor although there is a slight difference on the exact gap size\cite{LZhao2008XJZhou,HDing2008NLWang,MZHasan2008LWray,DVEvtushinsky2009SVBorisenko}. However, one laser-ARPES measurement on the optimally-doped (Ba$_{0.6}$K$_{0.4}$)Fe$_2$As$_2$ superconductor\cite{TShimojima2011SShin} gave a rather different result: similar momentum-dependence of the  superconducting gap is observed on all three hole pockets around $\Gamma$ with a maximum gap size of $\sim$3 meV (this gives a $2\Delta/k_BT_C$  value of 1.7 which is much smaller than the BCS value of 3.53 for an s-wave superconductor). Solution of this apparent disparity on the superconducting gap is important in proposing and examining theories to understand the superconductivity mechanism in the iron-based superconductors. Using the same laser photon energy (6.994 eV) as used in Ref.\cite{TShimojima2011SShin}, our laser-ARPES measurement results of the superconducting gap are rather different from that reported in Ref.\cite{TShimojima2011SShin} but fully consistent with all the other ARPES measurements\cite{LZhao2008XJZhou,HDing2008NLWang,MZHasan2008LWray,DVEvtushinsky2009SVBorisenko}. We note that the superconducting coherence peaks observed in our laser-ARPES measurements (Fig. 1g and 1h) are much sharper than that reported in Ref.\cite{TShimojima2011SShin} and also all other previous ARPES measurements\cite{LZhao2008XJZhou,HDing2008NLWang,MZHasan2008LWray,DVEvtushinsky2009SVBorisenko}.  Also within the same energy window, only single sharp peak is observed in our measurements (Fig. 1g and 1h) instead of two peaks observed in Ref.\cite{TShimojima2011SShin}. Our present results ask for reexamination of the results reported in Ref.\cite{TShimojima2011SShin}. The  precise determination of the gap size, symmetry and scattering rate also provides important information to understand superconductivity mechanism in (Ba$_{0.6}$K$_{0.4}$)Fe$_2$As$_2$ superconductor\cite{PJHirschfeld2011IIMazin,FWang2011DHLee}.

\vspace{3mm}
In order to further investigate the superconducting transition in the (Ba$_{0.6}$K$_{0.4}$)Fe$_2$As$_2$ superconductor,  we have carried out detailed temperature-dependent measurements, as shown in Fig. 2. Fig. 2a(1-8) show the photoemission images measured along the $\Gamma$- M direction (the location of the momentum cut is shown in the bottom-left of Fig. 2b) at different temperatures.  All the data are divided by the corresponding Fermi distribution function in order to show the band above the Fermi level, and highlight the opening of an energy gap around the Fermi level.  At low temperatures  below T$_c$$\sim$38.5 K (Fig. 2a(1-5)), a clear spectral weight suppression can be seen around the Fermi level for both the $\alpha$ and $\gamma$ bands which indicates the opening of the superconducting gap. Particle-hole symmetry is observed when the bands above the Fermi level are visible.  When the temperature reaches T$_c$  and above(Fig. 2a(6-8)) , the spectral weight around the Fermi level gets filled (Fig. 2a(1-5))  which indicates the closing of the superconducting gap. Fig. 2b and 2d shows EDCs measured at different temperatures for the two Fermi momenta of the cut crossing both $\alpha$ and $\gamma$ pockets, respectively. The corresponding symmetrized EDCs are shown in Fig. 2c and Fig. 2e.  Sharp superconducting coherence peaks develop at low temperature along both the inner (Fig. 2b) and outer pockets (Fig. 2d); the EDC width (full width at a half maximum) is $\sim$10 meV for the $\alpha$ band and $\sim$7 meV for the $\gamma$ band at 13 K.  With increasing temperature, the superconducting coherence peak gets weak and disappears above T$_c$.  Such a temperature evolution of EDCs is similar to that found in underdoped and optimally-doped cuprate superconductors near the antinodal region\cite{AVFedorov1999PDJohnson}.

We have found that the superconducting gap of the (Ba$_{0.6}$K$_{0.4}$)Fe$_2$As$_2$ superconductor exhibits an anomalous temperature dependence.  In order to get superconducting gap size at different temperatures on the two Fermi pockets, the symmetrized EDCs in Fig. 2c and 2e are fitted with the standard phenomenological model\cite{MRNorman1998JCCampuzano}.  The superconducting gap thus obtained is shown in Fig. 2f and the corresponding single-particle scattering rate is shown in Fig. 2g,  marked as Sample$\sharp$1.  The results have been reproduced by independently measuring many other samples; the results for the other two samples are also plotted in Fig. 2f and 2g (their original EDCs and fitting results can be found in Fig. S2). Surprisingly, the superconducting gap size is nearly constant below T$_c$ but drops to zero at T$_c$ abruptly, for both the inner $\alpha$ pocket and outer $\gamma$ pocket (Fig. 2f).   It deviates strongly from the BCS theory where the superconducting gap will gradually decrease with increasing temperature following a form shown as the dashed lines in Fig. 2f.   Deviation of the measured superconducting gap from the standard BCS form was also observed in (Ba,K)Fe$_2$As$_2$ before\cite{HDing2008NLWang}. In underdoped cuprate superconductors, it is observed that the superconducting gap shows little change with temperature in the superconducting state in the antinodal region. However, the gap does not close at T$_c$ but extends to a temperature T$^{*}$ that is much higher than T$_c$ associated with the pseudogap formation\cite{MRNorman1998JCCampuzano}. In the optimally-doped (Ba$_{0.6}$K$_{0.4}$)Fe$_2$As$_2$ superconductor, no indication of pseudogap formation is found because the superconducting gap drops to zero at T$_c$. At the same time, the single-particle scattering rate (Fig. 2g), extracted from fitting the symmetrized EDCs in Fig. 2c and 2e,  also shows a similar temperature dependence:  it shows little change with temperature in the superconducting state (Fig. 2g) . Above T$_c$, it becomes literally infinite because the symmetrized EDCs become nearly flat near the Fermi level with the disappearance of the coherence peak above T$_c$.

On the other hand, the spectral weight of the superconducting coherence peak shows a clear temperature dependence in the superconducting state, as seen in Fig. 2h.  To extract the spectral weight of the superconducting coherence peak, the symmetrized EDC in the superconducting state is plotted against that in the normal state, as shown in the inset of Fig. 2h. Here the peak generation area (pink region) is defined as  \emph{A} while the peak depletion area (green region) is defined as  \emph{B}. The variation of the spectral weights \emph{A} and \emph{B} with temperature, from three independent measurements on three samples, is shown in Fig. 2h.  Different from the temperature evolution of the gap size (Fig. 2f) and the scattering rate (Fig. 2g) , both the spectral weights \emph{A} and \emph{B} decrease monotonically with increasing temperature.  The superconducting spectral weight seems not drop to zero at T$_c$ but slightly above the superconducting transition temperature. This may be caused by either  superconducting fluctuation or possible difference of superconductivity between the sample surface and the bulk. The exact origin needs further investigation.  To facilitate comparison, we also plot the temperature dependence of the superfluid density measured in (Ba,K)Fe$_2$As$_2$ superconductor\cite{KHashimoto2009YMatsuda}. The variation of the superconducting coherence peak (generation area \emph{A}) shows a similar temperature dependence with the superfluid density in spite of the difference at T$_c$.  We note that the spectral weight \emph{A} is much larger than \emph{B} in Fig. 2h.  This indicates that the total spectral weight is not conserved near the Fermi level region when the sample enters superconducting state from the normal state: some extra spectral weight is generated in the superconducting state of the optimally-doped (Ba$_{0.6}$K$_{0.4}$)Fe$_2$As$_2$.  Such an acquisition of extra spectral weight in the superconducting state is also quite unusual; how to understand this breakdown of the spectral weight conservation needs further investigation.

It is interesting to find from Fig. 2c and 2e that, as soon as the sample enters the normal state above T$_c$, not only the superconducting coherence peak disappears, but also there is no quasiparticle peak in the normal state. The symmetrized EDCs in the normal state is basically flat near the Fermi level region for both the inner and outer Fermi pockets.   We also devide the  EDCs by Fermi distribution function at the corresponding temperature to remove the Fermi cutoff and still no peak feature is present around the Fermi level (Fig. S3). This indicates a fully incoherent normal state for the two Fermi pockets in the (Ba$_{0.6}$K$_{0.4}$)Fe$_2$As$_2$ superconductor. This is in a strong contrast to a usual Fermi liquid system where a well-defined quasiparticle peak is expected along the Fermi surface in the normal state.  To check the momentum dependence of the  non-Fermi liquid  behavior in the (Ba$_{0.6}$K$_{0.4}$)Fe$_2$As$_2$ superconductor, we show Fermi surface mapping and photoemission spectra along the Fermi surface at temperatures below and above T$_c$ in Fig. 3.  Two clear Fermi surface sheets are observed in Fig. 3a measured in the supercondcuting state and in Fig. 3b measured in the normal state. Fig. 3c and 3e shows the original EDCs measured at 13 K in the supercondcuting state (black lines) and at 50 K in the normal state (red lines) along the inner and outer pockets, respectively. The corresponding symmetrized EDCs are shown in Fig. 3d and 3f.  It is clear that in the superconducting state all the EDCs and symmetrized EDCs along the two Fermi surface sheets show very sharp superconducting coherence peaks (Fig. 3c-f).  In stark contrast, all the EDCs in the normal state do not show any peak feature around the Fermi level (Fig. 3c and 3e) and the corresponding symmetrized EDCs smoothly cross the Fermi level without any peaks (Fig. 3d and 3f).  Since we have covered nearly one quarter of both the Fermi surface sheets, considering the four-fold symmetry of the sample, our results indicate that the presence of sharp coherent peaks in the superconducting state and lack of quasiparticles in the normal state are universal for the entire Fermi surface sheets. We note that this absence of quasiparticles along the Fermi surface is not universal for all the iron-based superconductors; the quasiparticles are observed along the Fermi surface in the normal state in LiFeAs\cite{Borisenko2010PRL} and FeSe$_{0.5}$Te$_{0.5}$\cite{Zhangpeng2018Science} superconductors.

It is natural to ask whether the absence of quasiparticles is due to the photoemission matrix element effect. This is unlikely because under the same measurement condition, sharp superconducting coherence peak can be observed.  Also, the previous ARPES measurements under various measurement conditions reveal no signature of quasiparticles along the two Fermi surfaces\cite{LZhao2008XJZhou,HDing2008NLWang,MZHasan2008LWray,DVEvtushinsky2009SVBorisenko,YMXu2011HDing}.  Another possibility is whether this lack of quasiparticle in the normal state is due to strong impurity or disorder scattering. Disorder in a normal metal usually will result in enhancement of the electron scattering which can reduce the lifetime of quasiparticles\cite{EMiranda2005VDobrosavljevic}, thus broadening the EDC peak at k$_F$ near the Fermi level correspondingly.  From our transport measurement results (Fig. 1a), electron scattering from the impurity or disorder is nearly negligible as judged from the residual resistivity at the zero temperature. We also checked the sample ageing effect by cooling the sample to low temperature again after warming up above T$_c$.  The reemergence of the sharp superconducting coherence peak rules out the possibility of sample ageing for the absence of the quasiparticle peak in the normal state (Fig. S4).  The present results have provided direct spectroscopic evidence that the absence of quasiparticle peaks in the normal state is intrinsic, and the normal state of the optimally-doped (Ba$_{0.6}$K$_{0.4}$)Fe$_2$As$_2$  superconductor is a non-Fermi liquid. We note that EDC peaks can be observed near M point in the normal state of the optimally-doped (Ba$_{0.6}$K$_{0.4}$)Fe$_2$As$_2$ superconductor\cite{LZhao2008XJZhou}, but this does not alter the conclusion that this is a non-Fermi liquid system because in a Fermi liquid system every Fermi surface should show quasiparticle peaks in the normal state.

The non-Fermi liquid behaviors in the normal state of (Ba$_{0.6}$K$_{0.4}$)Fe$_2$As$_2$ superconductor can be further examined by the photoemission lineshape and electron self-energy analyses, as shown in Fig. 4.   Fig. 4e and 4f show photoemission images measured at 13 K in the supercondcuting state and  50 K in the normal state, respectively, taken along the momentum cut marked as a black line in Fig. 4a.  Fig. 4d shows two EDCs measured at 13 K and 50 K at the Fermi momentum for the $\alpha$ band. Fig. 4b and 4c show MDCs  at the Fermi level at 13 K from Fig. 4e and 50 K from Fig. 4f, respectively.  By fitting MDCs at different binding energies, dispersions for the $\alpha$  and $\gamma$ bands in the superconducting state (13 K) and normal state (50 K) are shown in Fig. 4g and the corresponding MDC width as a function  of energy is shown in Fig. 4h.   A clear kink is developed at $\sim$18 meV in the $\alpha$ band dispersion at 13 K (Fig. 4g).   By assuming an empirical linear bare band (dashed line in Fig. 4g) for the $\alpha$ band dispersion, the effective real part of electron self-energy is obtained , as shown in the inset of Fig. 4g. A clear peak at $\sim$18 meV  shows up in the 13 K data which is consistent with the kink in the dispersion.  Different from the $\alpha$ band, the $\gamma$ band does not show clear kink feature which is possibly due to the orbital selectivity of electron-boson coupling\cite{HDing2009PRichard}.  Fig. 4i shows the MDCs at the Fermi level along different Fermi surface angles (defined in the bottom-left figure in Fig. 4j). All the MDCs are fitted with two Lorentzians and the fitting MDC widths are shown in Fig. 4j. It is nearly isotropic along the Fermi surface and similar in value for both $\alpha$ and $\gamma$ Fermi surface sheets. The images from the momentum cuts along various angles and their MDC-derived dispersions are also obtained, as shown  in Fig. S5 (Supplementary Materials).  Similar to the MDC width, all the dispersions along different angles nearly overlap on top of each other showing high isotropy around the Fermi surface for both Fermi surface sheets.  Based on the dispersions measured in the normal state (50 K) in Figs. 4g and S5, the Fermi velocities along the two Fermi surfaces are obtained as shown in Fig. 4k. Two sets of nearly isotropic Fermi velocity are observed: the inner $\alpha$ Fermi pocket possesses  a larger Fermi velocity than the outer $\gamma$ Fermi pocket. Fig. 4l shows the product of the MDC width at the Fermi level in Fig. 4j and the Fermi velocity in Fig. 4k for the two Fermi surface sheets.

The MDC-EDC dichotomy in Fig. 4 further corroborates the non-Fermi liquid behaviors in the normal state of (Ba$_{0.6}$K$_{0.4}$)Fe$_2$As$_2$ superconductor. From ARPES measurement, usually the electron scattering rate can be obtained in two ways: either the linewidth of EDC, or the MDC width at the Fermi level multiplied by the Fermi velocity.  In the Fermi liquid system, these two ways should give the same results.  However, in the (Ba$_{0.6}$K$_{0.4}$)Fe$_2$As$_2$ superconductor, the EDCs in the normal state show no sign of peaks (Fig. 3 and Fig. 4d), but the MDCs in the normal state show clear peaks (Fig. 4i and 4c). Literally speaking, the absence of EDC peak indicates the scattering rate is infinite, while the MDC peak, together with the Fermi velocity, gives a finite ``MDC" scattering rate along the two Fermi surface sheets (Fig. 4l). This MDC-EDC dichotomy clearly signals the breakdown of the Fermi liquid picture in the (Ba$_{0.6}$K$_{0.4}$)Fe$_2$As$_2$ superconductor. The nearly constant MDC width as a function of energy for the inner $\alpha$ band in the normal state (Fig. 4h) also points to the non-Fermi liquid behavior since it is expected to be quadratic in the Fermi liquid system.

Our results have established that the optimally-doped (Ba$_{0.6}$K$_{0.4}$)Fe$_2$As$_2$ superconductor is a non-Fermi liquid system with a well-defined Fermi surface but without well-defined quasiparticles along the Fermi surfaces near $\Gamma$ in the normal state.  This is apparently different from conventional superconductors which have both well-defined Fermi surface and well-defined quasiparticles along the Fermi surface.  This is also different from the underdoped cuprate superconductors which exhibit non-Fermi liquid normal state. First, near the antinodal region of the underdoped cuprate superconductors, there is neither well-defined Fermi surface due to the pseudogap formation nor existence of quasiparticles in the normal state\cite{AGLoeser1996AKapitulnik,HDing1996JGiapintzakis}. Second, the electronic structure of the cuprate superconductors is highly anisotropic where the nodal and antinodal regions exhibit distinct behaviors\cite{XJZhouPRL,KShenScience}. The strange metal behavior of the normal state is closely related to this momentum-dependent electronic structure.   However, in optimally-doped (Ba$_{0.6}$K$_{0.4}$)Fe$_2$As$_2$ superconductor, the electronic structure along the two Fermi surface sheets near $\Gamma$ is highly isotropic.  In the underdoped cuprate superconductors, it is known that the superconducting gap around antinodal region exhibits an anomalous behaviour by keeping nearly constant with increasing temperature in the superconducting state, smoothly crossing the superconducting transition, and persisting to a temperature much higher than the superconducting transition temperature\cite{ZXShen2007WSLee,XJZhou2009JMeng2}. In the optimally-doped (Ba$_{0.6}$K$_{0.4}$)Fe$_2$As$_2$ superconductor, the unusual temperature dependence of the superconducting gap (Fig. 2f), the non-conservation of spectral weight between the normal and superconducting states (Fig. 2h),  may be closely related to the non-Fermi liquid nature of the system.

The absence of quasiparticles along the Fermi surface points to a fully incoherent normal state in the optimally-doped (Ba$_{0.6}$K$_{0.4}$)Fe$_2$As$_2$ superconductor. The microscopic origin of the non-Fermi liquid behaviors is critical to understand its superconductivity. The first possibility is the existence of extremely strong source of electron scatterers. Although we believe the impurity or disorder is unlikely to account for the absence of quasiparticles because of high quality and cleanness of our samples, it is important to examine on the effect of electron-electron interaction, and electron scattering from bosons like spin fluctuations\cite{PCDai2015RMP}.  Another possibility is the existence of quantum critical point in the optimally-doped region of iron-based superconductors.  This has been reported in the optimally-doped BaFe$_2$(As,P)$_2$ superconductors\cite{YMatsuda2014TShibauchi,JGAnalytis2014IRFisher,HHKuo2016IRFisher,YGu2017SLi}. The strong quantum fluctuations could destroy the long-lived quasiparticles and result in non-Fermi liquid behavior in the normal state.  However, whether there exists a quantum critical point in (Ba$_{1-x}$K$_{x}$)Fe$_2$As$_2$ system needs further investigations.   The quasiparticles can also be extincted due to the closeness to the orbital-selective Mott phase because of the interplay between Hund's coupling and electron-electron Coulomb interaction\cite{Ryu2013QMSi,SAcharya2018ATaraphder}.  In this case,  the quasiparticle spectral weight for some  orbitals will be heavily suppressed. However, the absence of quasiparticles in (Ba$_{0.6}$K$_{0.4}$)Fe$_2$As$_2$ seems to be momentum and Fermi surface independent around the $\Gamma$ Point. We note that the resistivity of Ba$_{0.6}$K$_{0.4}$)Fe$_2$As$_2$ superconductor is metallic in the normal state (Fig. 1a) and there is a Drude-like peak in its optical conductivity in the normal state\cite{GLi2008NLWang,BXu2014XGQiu}. More efforts are needed to understand the origin of the absence of quasipaticles along Fermi surface, its effect on the normal state of (Ba$_{0.6}$K$_{0.4}$)Fe$_2$As$_2$ superconductor and its associated unusual behaviors like the exotic temperature evolution of the superconducting  gap.

\vspace{3mm}

\vspace{3mm}

\section{Summary}

In summary, we have investigated the electronic structure and superconducting gap of the optimally-doped (Ba$_{0.6}$K$_{0.4}$)Fe$_2$As$_2$ superconductor both in the superconducting state and in the normal state with our high-resolution laser-based ARPES system with unprecedent precision. In the superconducting state, we observed Fermi surface dependent superconducting gap and its extremely isotropic momentum dependence. This has solved a long-standing controversy concerning the superconducting gap of the optimally-doped (Ba$_{0.6}$K$_{0.4}$)Fe$_2$As$_2$ superconductor. We found that the superconducting gap exhibits an unusual temperature dependence and there is an extra gain of spectral weight when the sample enters the superconducting state.   Sharp superconducting coherence peaks are observed in the superconducting state, but no signature of quasiparticle peaks is observed in the normal state. Together with the photoemission lineshape and self-energy analyses, we have provided direct spectroscopic evidence that the optimally-doped (Ba$_{0.6}$K$_{0.4}$)Fe$_2$As$_2$ superconductor is a non-Fermi liquid system. Moreover,  we have established optimally-doped (Ba$_{0.6}$K$_{0.4}$)Fe$_2$As$_2$ superconductor as a new system which has well-defined Fermi surface but without well-defined normal state quasiparticles along the Fermi surface. This makes it distinct from the conventional superconductors with both well-defined Fermi surface and well-defined normal state quasiparticles, and the underdoped cuprate superconductors with no well-defined Fermi surface and no well-defined normal state quasipartlces near the antinodal region.   These results  will provide significant insights to understand the electronic properties of the ``abnormal" normal state and put strong constraints on theories in understanding the superconductivity mechanism of unconventional superconductors.

\vspace{3mm}

$^{*}$Corresponding author:  LZhao@iphy.ac.cn,  XJZhou@iphy.ac.cn.

\vspace{3mm}

\noindent {\bf Acknowledgement}\\
We thank Dunghai Lee and Qijin Chen  for discussions. This work is supported by the National Key Research and Development Program of China (Grant No. 2016YFA0300300) and 2017YFA0302900, the Strategic Priority Research Program (B) of the Chinese Academy of Sciences (Grant No. XDB07020300, XDB25000000), the National Basic Research Program of China (Grant No. 2015CB921000), the National Natural Science Foundation of China (Grant No. 11334010), and the Youth Innovation Promotion Association of CAS (Grant No. 2017013).

\vspace{3mm}

\noindent {\bf Author Contributions}\\
 X.J.Z., L.Z. and J.W.H. proposed and designed the research. J.W.H. contributed in sample growth. J.W.H., L.Z., C. L.,  Q.G., J.L., Y.H., Y.X., Y.Q.C., D.S.W., Y.D., C.H., G.D.L., Q.Y.W., S.J.Z., Z.M.W., F.F.Z., F.Y., Q,J,P., Z.Y.X., C.T.C. and X.J.Z. contributed to the development and maintenance of the Laser-ARToF system and related software development. H.X.Z. and X.L.D. contributed to the magnetic measurement. J.W.H. carried out the ARPES experiment with Q.G., L.Z. and C.L.. J.W.H., L.Z. and X.J.Z. analyzed the data. J.W.H.,  L.Z. and X.J.Z. wrote the paper. All authors participated in discussion and comment on the paper.\\

\vspace{3mm}

\noindent{\bf Additional information}\\
Supplementary information is available in the online version of the paper.
Correspondence and requests for materials should be addressed to  L.Z. and X.J.Z.

\newpage

\begin{figure*}[tbp]
\begin{center}
\includegraphics[width=0.8\columnwidth,angle=0]{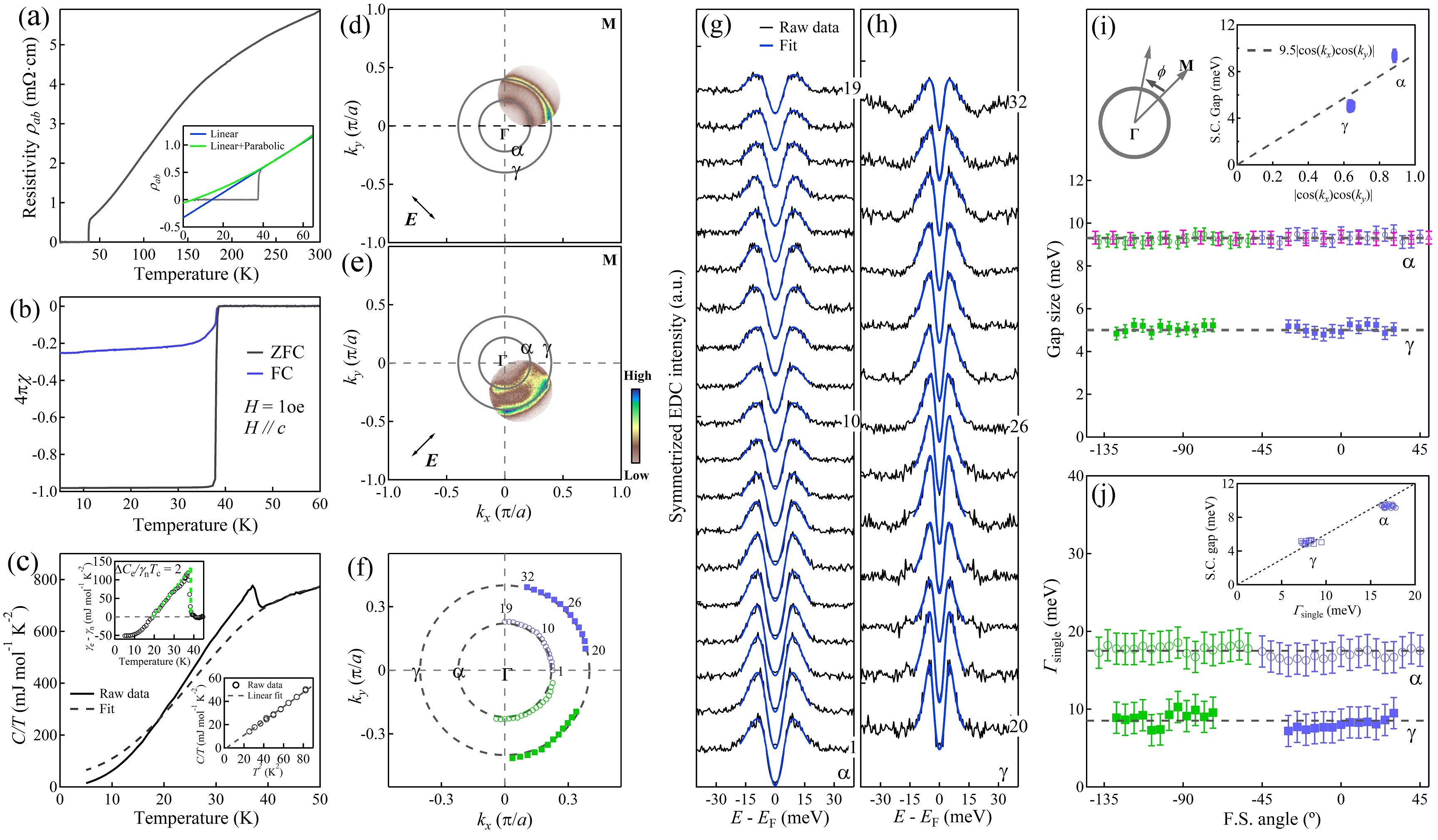}
\end{center}

\caption{{\bf Sample Characterization and Momentum Dependence of the Superconducting Gap  in Optimally-Doped (Ba$_{0.6}$K$_{0.4}$)Fe$_2$As$_2$ Superconductor.} (a) Resistivity measurement within $\it ab$-plane. The onset superconducting transition temperature for this sample is 39 K. The bottom-right inset shows linear and linear plus quadratic fitting of the low temperature resistivity. (b) Magnetic susceptibility measurement with zero-field-cooling (ZFC) and field-cooling (FC) methods. It exhibits a sharp transition at 38.5 K with a narrow transition width of  $\sim$0.5 K. (c) Specific heat coefficient $\gamma$ = C/T as a function of temperature (T).  The dashed line shows the normal-state specific heat that is obtained by fitting using the method reported before\cite{GMu2009HHWen} while keeping the entropy conservation. The up-left inset shows electronic specific heat contribution with the normal-state part subtracted. The bottom-right inset shows temperature dependent specific heat in C/T vs T$^{2}$. The dashed line represents a linear fit that gives a zero residual value at zero temperature. (d) Fermi surface mapping  at  13 K.  (e) Same as (d) but with the sample rotated by 90 degrees with respect to that in (d). (f) Fermi surface obtained from (d) and (e). (g) Symmetrized EDCs along the inner hole pocket  $\alpha$; the corresponding location of momentum is marked in (f).  The blue solid lines overlaid on the raw EDCs are fitting results. (h) Same as (g) for the outer hole pocket $\gamma$. (i) Superconducting gap as a function of the Fermi surface angle defined in the top-left inset. The up-right inset plots superconducting gap size as a function of $\mid$$\cos(k_x)\cos(k_y)$$\mid$. The dashed line shows the best fit. The data points in magenta color are from another measurement as shown in Fig. S1 in the Supplementary Materials. (j)  Single-particle scattering rate obtained by fitting the symmetrized EDCs in (g) and (h) using a phenomenological model as a function of Fermi surface angle. Inset plots the corresponding superconducting gap size as a function of the scattering rate. The dashed linear line is a guide to the eye.
}
\end{figure*}

\begin{figure*}[tbp]
\begin{center}
\includegraphics[width=1.0\columnwidth,angle=0]{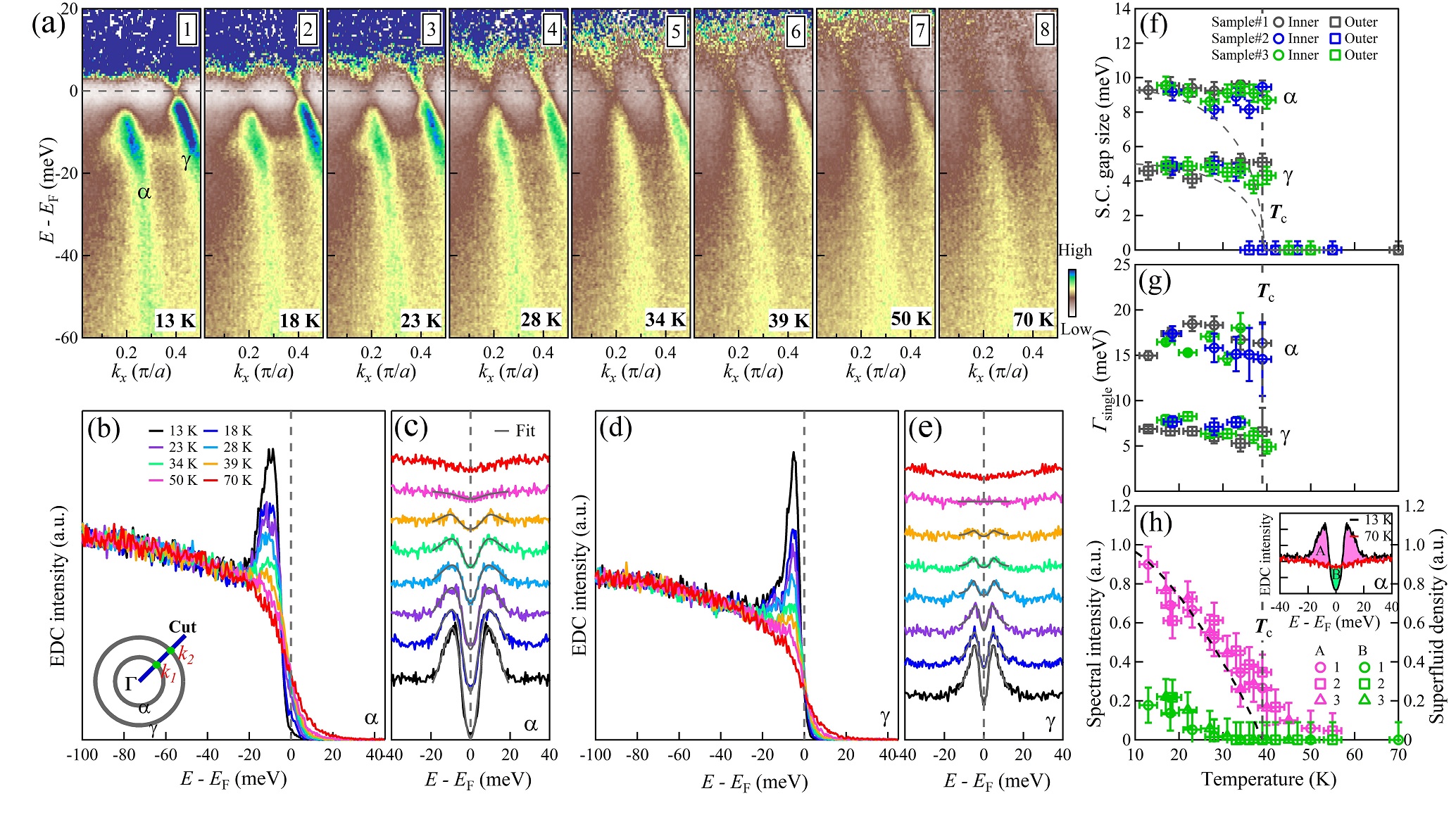}
\end{center}

\caption{{\bf Temperature Dependence of Band Structure and Superconducting Gap for the Optimally-Doped (Ba$_{0.6}$K$_{0.4}$)Fe$_2$As$_2$ (T$_c$=38.5 K).} (a1-a8) Band structure measured at different temperatures along the momentum cut marked by the blue line in the bottom-left inset of (b). The observed two bands are labelled in (a1) as $\alpha$ band for the inner hole pocket and $\gamma$ for the outer hole pocket.   (b) EDCs measured at different temperatures for the Fermi momentum k$_1$ on the inner Fermi pocket.   (c) Corresponding symmetrized EDCs of (b) with fitting results overlaid on the raw data. The curves are offset for clarity.  (d-e) Same as (b-c) for the Fermi momentum k$_2$ on the outer Fermi pocket.  (f) Temperature dependence of the  superconducting gap for the inner and outer Fermi pockets measured on different samples. The empty circles (squares) represent  data for the inner (outer) hole pocket. (g) The single-particle scattering rate as a function of temperature obtained by fitting the symmetrized EDCs in (c) and (e). (h) Integrated spectral intensity of the superconducting coherent peak (shaded pink area in the inset) and superconducting coherent dip (shaded green area in the inset) as a function of temperature. The black dashed line is superfluid density of (Ba,K)Fe$_2$As$_2$  extracted from\cite{KHashimoto2009YMatsuda}.
}
\end{figure*}

\begin{figure*}[tbp]
\begin{center}
\includegraphics[width=1.0\columnwidth,angle=0]{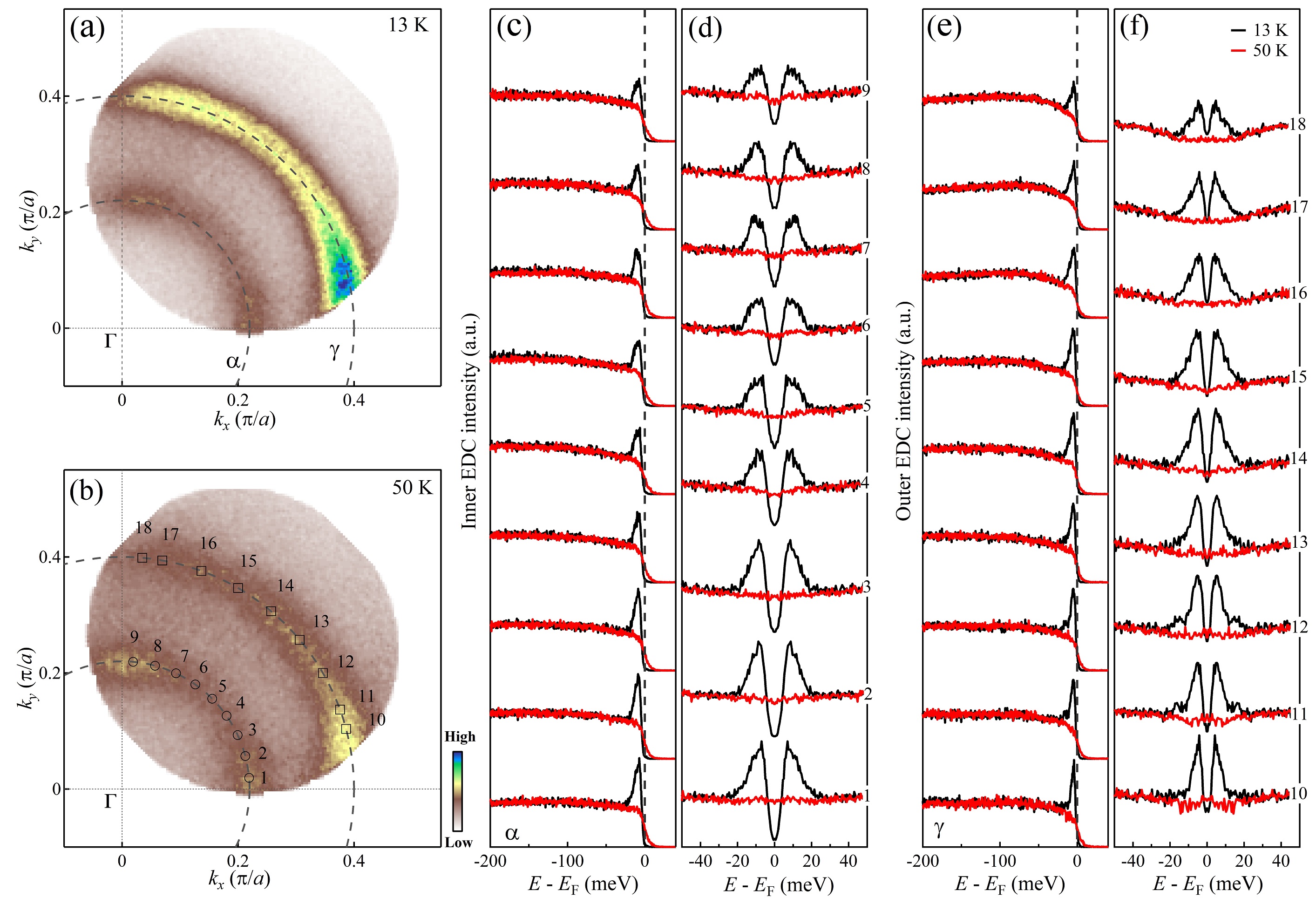}
\end{center}

\caption{{\bf Photoemission Spectra on the Two Fermi Surface Sheets in the Normal and Superconducting States for the (Ba$_{0.6}$K$_{0.4}$)Fe$_2$As$_2$ (T$_c$=38.5 K) Superconductor.} (a) Fermi surface mapping at 13 K in the superconducting state.  (b) Fermi surface mapping at 50 K in normal state.  (c) EDCs  along the inner hole-like $\alpha$ Fermi  pocket measured at 13 K (black line) and 50 K (red line). The location of the Fermi momenta is marked in (b) by empty circles. The EDCs are offset along vertical axis for clarity.  (d) Corresponding symmetrized EDCs of (c).  (e) EDCs  along the outer hole-like $\gamma$ Fermi  pocket measured at 13 K (black line) and 50 K (red line). The location of the Fermi momenta is marked in (b) by empty squares. The EDCs are offset along vertical axis for clarity.  (f) Corresponding symmetrized EDCs of (e).  Flat symmetrized EDCs near the Fermi level at 50K indicate the absence of coherent quasiparticles in the normal state.
   }
\end{figure*}

\begin{figure*}[tbp]
\begin{center}
\includegraphics[width=1.0\columnwidth,angle=0]{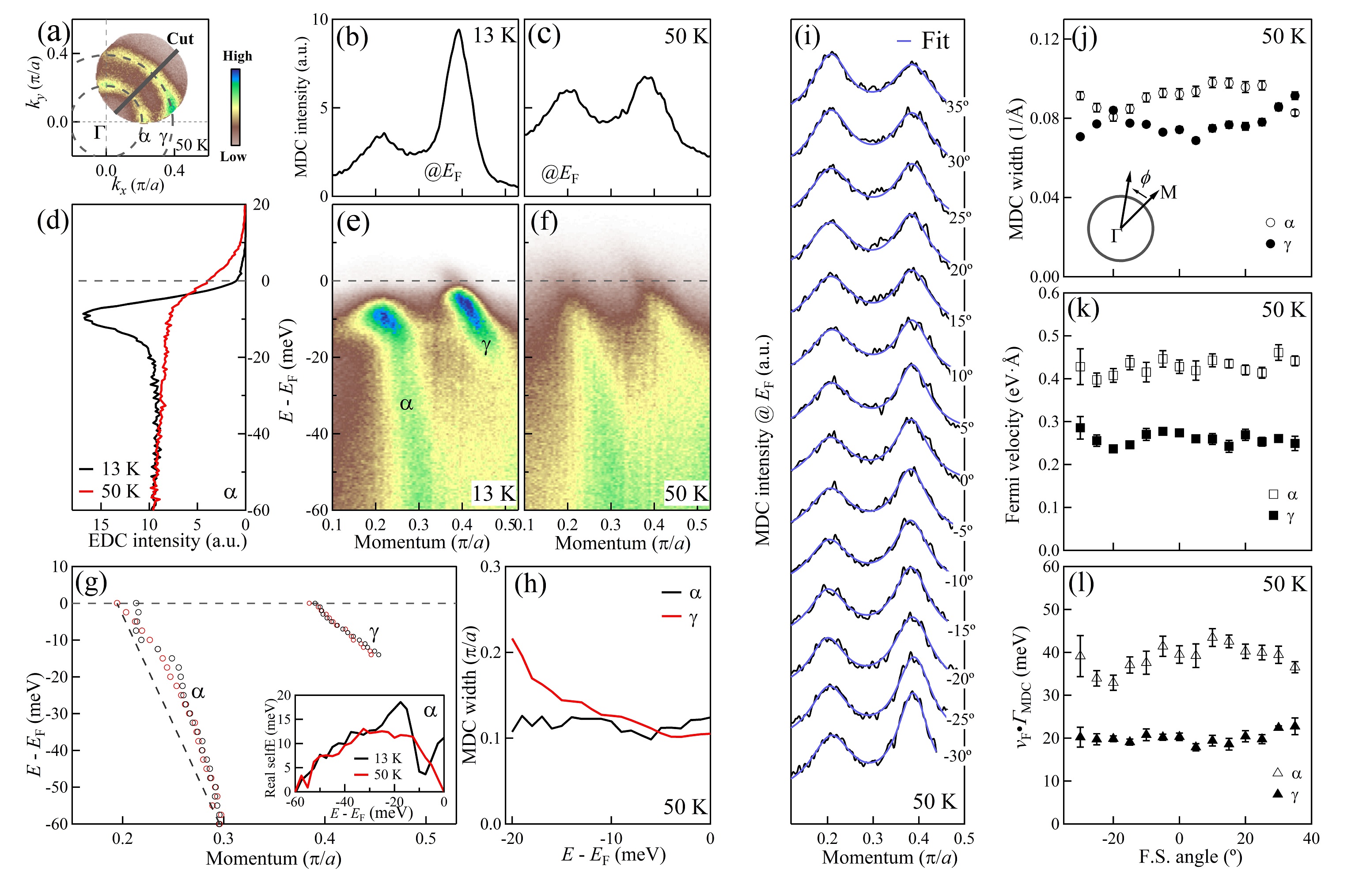}
\end{center}

\caption{{\bf Non-Fermi Liquid Electronic Behaviors in the Normal State of Optimally-Doped (Ba$_{0.6}$K$_{0.4}$)Fe$_2$As$_2$ Superconductor.} (a) Fermi surface mapping at 50K.  (b) MDC at E$_F$ for the momentum cut indicated by the black line  in (a) at 13 K. The cutting is along $\Gamma$-M high symmetry direction.  (c) Same as (b) at 50 K. (d) EDCs at k$_F$ of the $\alpha$ band at 13 K (black) and 50 K (red). (e) Band structure along the momentum cut indicated by the black line in (a) at 13K.  (f) Same as (e) at 50 K. (g) Band dispersions obtained by MDC fitting of (e) and (f) with two Lorentzian peaks. The black dashed line is an empirical linear bare band. Inset shows the real part of the electron self-energy at 13 K and 50 K. (h) MDC width of the fitting results for the $\alpha$ band and $\gamma$ band in (f) as a function of energy. (i) MDCs at E$_F$ along momentum cuts  with different Fermi surface angles as indicated in the inset of (j).  The MDCs are overlaid with fitted curves with two Lorentzians. The MDCs are offset along the vertical axis for clarity.  (j) MDC width along the two hole pockets plotted as a function of the Fermi surface angle. (k) Fermi velocity v$_F$ of the two hole pockets plotted as a function of the Fermi surface angle. (l) The product of the MDC width and Fermi velocity, v$_F\cdot\Gamma_{MDC}$,  of the two hole pockets plotted as a function of the Fermi surface angle.
   }

\end{figure*}

\end{document}